\PassOptionsToPackage{utf8}{inputenc}
\documentclass[a4paper,12pt]{article}
\usepackage[colorlinks=false]{hyperref}
\usepackage{url}
\usepackage{comment}
\usepackage{caption}
\usepackage{subcaption}
\usepackage{graphicx}
\usepackage[margin=1.5cm]{geometry}
\usepackage{appendix}
\usepackage{longtable}
\usepackage{booktabs}
\usepackage{rotating}
\usepackage{hyperref}
\usepackage{array}
\usepackage{xcolor}
\usepackage{amsmath}
\usepackage{authblk}
\usepackage{float}
\usepackage{graphicx}
\usepackage{placeins}
\usepackage{xr}
\usepackage[numbers]{natbib} 

\newcounter{suppfigure}

\newcounter{supptable}
\renewcommand{\thesupptable}{S\arabic{supptable}}

\title{JINet: easy and secure private data analysis for everyone}
\author[1]{Giada Lalli} 
\author[2]{James Collier}
\author[3]{Yves Moreau}
\author[3]{Daniele Raimondi}
\affil[1]{BIO3 – Systems Medicine Lab, KU Leuven, Belgium}
\affil[2]{VIB, Belgium}
\affil[3]{ESAT-STADIUS, KU Leuven, Belgium}

\begin{document}
\maketitle

\abstract{\textbf{Summary:} JINet is a web browser-based platform intended to democratise access to advanced clinical and genomic data analysis software. It hosts numerous data analysis applications that are run in the safety of each User's web browser, without the data ever leaving their machine. JINet promotes collaboration, standardisation and reproducibility by sharing scripts rather than data and creating a self-sustaining community around it in which Users and data analysis tools developers interact thanks to JINets interoperability primitives. \\
\textbf{Availability:} The source code for JINet is available from our Git Repository: \url{https://github.com/GiadaLalli/JINet}. The JINet application is hosted at \url{https://jinet.thecolliers.xyz} \\
\textbf{Contact:} giada.lalli@kuleuven.be, james.collier@vib.be}

\maketitle
\section*{Introduction}
Advancing healthcare research, fostering innovation, and enhancing reproducibility and transparency in science requires comprehensive access to datasets and effective analysis tools, as well as data privacy and security measures. In healthcare, these elements are crucial for uncovering disease mechanisms, developing personalised treatments, and promoting collaborative problem-solving.

Currently, significant barriers hinder these goals. Limited access to datasets and inadequate analysis tools impede meaningful insights. Concerns over data privacy and security discourage collaboration and hamper data sharing. Additionally, interoperability challenges and disparate data formats complicate seamless integration and comprehensive analysis. Technical barriers and an exclusive research environment further restrict accessibility, especially for non-experts.

Various solutions have been proposed to overcome these challenges. Centralised repositories, workflows and cloud-based platforms, such as blockchain-based solutions \cite{jianjun2020research, wang2018medical}, Galaxy~\cite{galaxyproject}, Taverna \cite{Taverna}, Nextflow \cite{Nextflow}, and Huggingface~\cite{Huggingface}, offer avenues for collaboratively storing, processing, and sharing large datasets, typically involving commitments such as funding, time, and trust. However, these solutions have notable limitations that impact their effectiveness and widespread adoption. For instance, Taverna relies heavily on third-party web services, which can be unstable; disruptions in these third-party services can lead to execution failures. Moreover, its workflows tend to grow complex due to the integration of numerous adaptive modules for data format transformations, and the platform's support for virtualization is limited.
Similarly, Galaxy requires specific software and reference datasets to be locally available, increasing technical complexity. The platform also faces challenges in recording and exporting provenance information due to the absence of a standardised schema or vocabulary.
Nextflow, while flexible, demands some technical skill from its users. Users must have some understanding of programming, and requires them to manually specify software and hardware dependencies. This lack of standardisation in execution and specification complicates exploring and utilising provenance information \cite{cohen2024workflows}. Thus, these solutions often fall short due to data security, patient privacy, usability, and regulatory compliance concerns that are difficult to address  ~\cite{Dove2015}. Interoperability issues across different systems and data formats also hamper effective dataset utilisation. Moreover, existing platforms require significant computational resources, user training, and infrastructure, which may not be universally accessible to researchers, limiting widespread adoption and efficacy. Consequently, these shortcomings limit the effectiveness and widespread adoption of data-sharing-based solutions.

To address these challenges, we introduce JINet, a novel platform designed for data analysis and result sharing, especially within genomic and healthcare domains. JINet provides a portfolio of self-contained applications that allow data to be analysed locally, ensuring the security and privacy of sensitive information. Unlike Galaxy, Taverna, and Nextflow, which typically require data to be uploaded to a central server or cloud-based system, JINet runs analyses locally, removing the dependency on external web services and ensuring more consistent performance. Transparently to the user, JINet locally downloads the analysis script selected by the user and runs it within the sandbox provided by the user's web browser (e.g., Firefox, Chromium). In doing so, JINet removes the privacy concerns due to the risks of transferring data via potentially untrustworthy channels, by removing the need to transfer data \emph{tout court}, sharing analysis scripts instead.

JINet aims at developing a self-sustaining community around it, composed of Application Developers, Users, and Data Providers. Developers are incentivized to integrate their latest machine Learning and Data Analysis methods into JINet, making them readily available to as many users as possible. Conversely, users are attracted by the platform's easy access to the latest tools. To make this ecosystem self-sustainable, we ensured that integrating new analysis scripts into JINet is straightforward: with a simple permission request and script submission, they are standardised by JINet for automatic access without installation requirements.

JINet provides an intuitive environment for analysing complex healthcare datasets, empowering researchers to work with structured and unstructured data from diverse sources without compromising confidentiality. As a self-contained platform, JINet democratises access to analysis pipelines by encapsulating their inner complexity, allowing Users with any degree of IT proficiency to run complex analyses.
Moreover, JINet facilitates collaboration by enabling Users and Data Providers to share results and insights without compromising data privacy.

\begin{figure}[ht]
  \centering
  \includegraphics[scale=1]{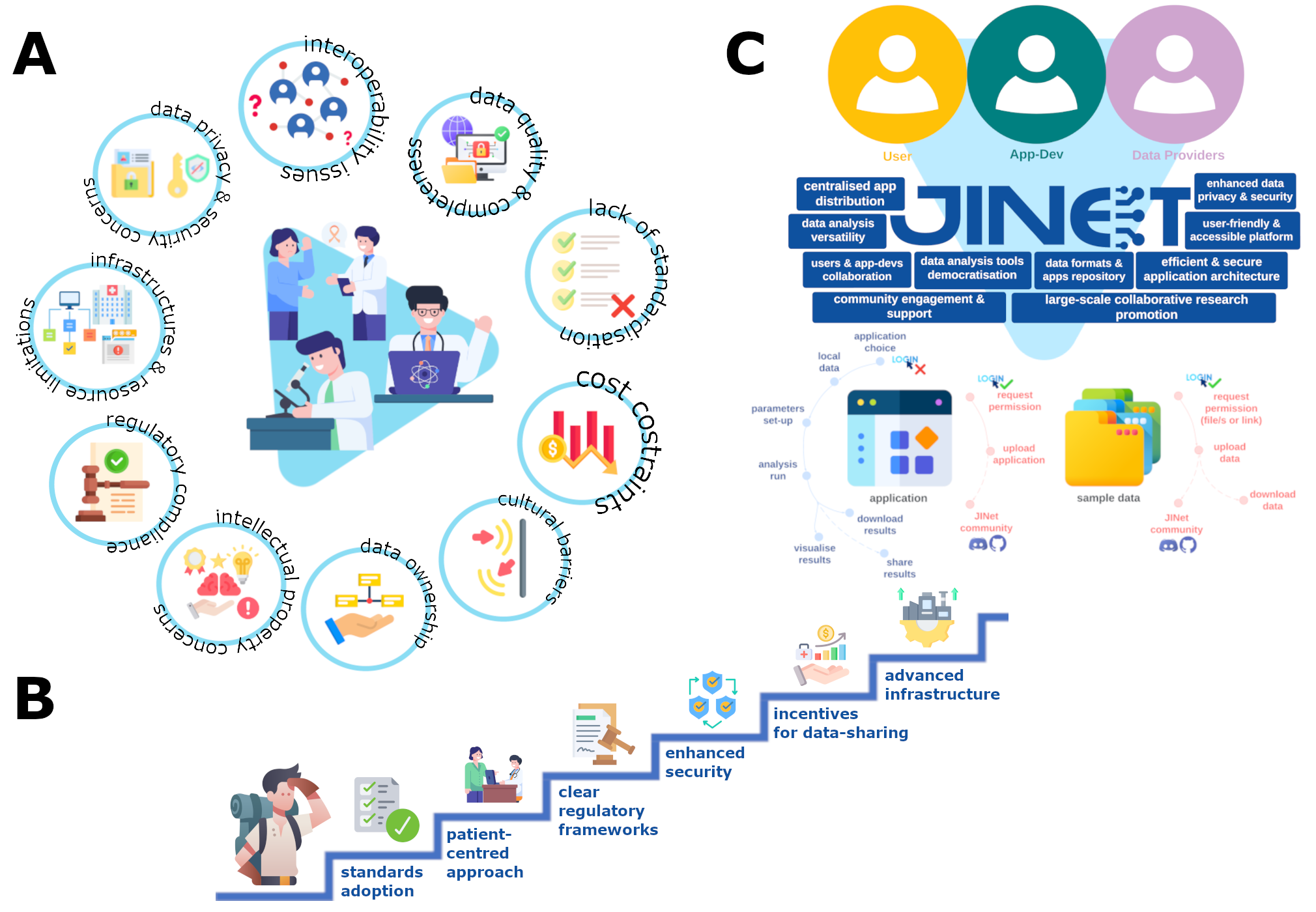}
  \caption{\textbf{A}: a figurative representation of the current major burdens encountered in data sharing within medical research. \textbf{B}: a strategic roadmap designed to address the challenges in data sharing and analysis for medical research, creating an environment conducive to robust and secure data-sharing practices. \textbf{C}: JINet ecosystem, facilitating collaboration among users, application developers, and data providers. JINet offers a centralised application distribution platform that enhances data privacy and security, supports versatile data analysis, and promotes community engagement and large-scale collaborative research. The ecosystem includes features such as data analysis tools democratisation, efficient and secure application architecture, and a repository for data formats and applications. The workflow shows the process from application choice to results visualisation and sharing, emphasising ease of use and security in data handling.}
  \label{fig:01}
\end{figure}

As illustrated in Figure~\ref{fig:01}C, JINet distinguishes between Users, Application Developers (app-devs), and Data Providers. Users, typically non-experts in programming, can run analyses without having to deal with the intrinsic complexity of the application they selected and without even logging in to JINet. Users browse the list of available applications, select the desired application, and then configure the application parameters. Application Developers can develop and make applications available to regular JINet users. Data Providers are entities that voluntarily share publicly available datasets for public use, such as benchmark datasets or examples of structured data. This kind of data might include datasets used in challenges, like the Dreamdata \cite{Dreamdata}, CAFA \cite{CAFA}, or Kaggle~\cite{kaggle}, or sample data from specific applications.


\FloatBarrier
\section*{Methods}

JINet is structured as two major components: a server and browser-based application runner. The server is an index of available applications and sample data. The client, running in a web browser, executes applications on user data without transmitting these data to the server or any other party.




\subsection*{JINet components}
\paragraph{Application Runner.} All JINet applications run inside a web browser sandbox on the user's machine. Three runtime environments are available for Python, R, and Javascript applications. Python and R interpreters are compiled to Web Assembly. The Python interpreter is provided by Pyodide \cite{pyodide_2021}, and the R interpreter is provided by WebR \cite{webR_2023}. This allows JINet to safely quarantine all data provided by users within the sandbox provided by the web browser.
All network traffic to remote servers, including the JINet application distribution server, is encrypted with SSL. Outgoing requests are vetted by the browser Content Security Policy effectively preventing applications from leaking private user-provided data.

\paragraph{Server.} The server component of JINet is responsible for authenticating users, distributing applications, and storing encrypted results (derived data) and sample data. Users are required to authenticate to publish an application or share results. Running an application does not require a login. The primary purpose of the application distribution server is to serve as a centralised hub and index of available applications. When a user selects an application to run, the application's source code is transferred to the user's web browser, where it is executed. The application distribution server has no knowledge of when, how, or with which parameters a user has executed the application.

The user may decide to store the results of an application (e.g., a plot or an output file) on their computer. They may also choose to share the results using JINet. JINet provides a mechanism for sharing results by encrypting them in the web browser with a key derived from a passphrase and storing them temporarily on the server.

\subsection*{Security \& Privacy}
While making scripts easy to use for non-experts, there are 2 negative security outcomes that JINet effectively prevents: 1) leaking user data to the internet, 2) accessing local user data without explicit permission.

To avoid these negative outcomes, JINet runs all applications within the web browser sandbox on the user's local machine. This means that user data is never transferred over the network to run an application. This sandbox is enforced by a locked-down content security policy that whitelists only a few required URLs that the application is allowed to connect to:

\begin{itemize}

    \item \url{https://jinet.thecolliers.xyz} the JINet application URL,
    \item \url{https://*.r-wasm.org} for the R interpreter and pre-compiled packages,
    \item \url{https://cdn.jsdelivr.net} for the Python interpreter and application styling,
    \item \url{https://pypi.org} for the Python package installer,
    \item \url{https://files.pythonhosted.org} for Python packages, and
    \item \url{https://raw.githubusercontent.com} for hosted application sources.
\end{itemize}

Importantly, this minimises the risks of adversarial or poorly written applications leaking private user data by transferring those data over the internet to a server that they control. The web browser sandbox also effectively prevents applications from arbitrarily accessing files on the user's local file system. Any local file access must be explicitly requested and allowed by the user before applications are given access to local files.

\subsubsection*{Sharing results}
Users may share results by transferring encrypted data to the JINet application distribution server. Data is encrypted in the browser client before transmission using a passphrase-derived key and the browser-provided implementation of AES-256 in CBC mode. Before encryption, JINet computes a SHA-256 checksum of the data to be sent. When the receiver decrypts the shared results, this checksum is verified to ensure that the results have not been tampered with.

\subsection*{Submitting an Application}
To submit an application to JINet, developers must write a script that defines a single entry-point function. This function serves as the primary interface for receiving and processing all external inputs, acting as the main gateway for executing the core logic. The types of arguments that the entry-point function can accept are limited by what JINet can display user-friendly interface controls for: filesystem paths, strings, integers, floats, and booleans. Developers may download dependencies and call into other functions from this entry point. The authors expect developers to set up the necessary computational environment (e.g., read input files from the filesystem) and then execute the main script functionality from the entry-point function. Due to parameter type restrictions, complex data structures such as data frames cannot be directly passed; instead, developers should accept filenames and read the data from these files.

The entry-point function must return either an HTML string or a filename. In the case that HTML is returned, it is embedded in the web browser interface directly for display to the user. Instead, if a filename is returned, a save button is displayed, allowing users to save the resulting file.

By following these guidelines, developers can trivially integrate analysis scripts into JINet, facilitating seamless data analysis without the risks of direct data sharing.

\begin{figure}[ht]
  \centering
  \includegraphics[width=\textwidth]{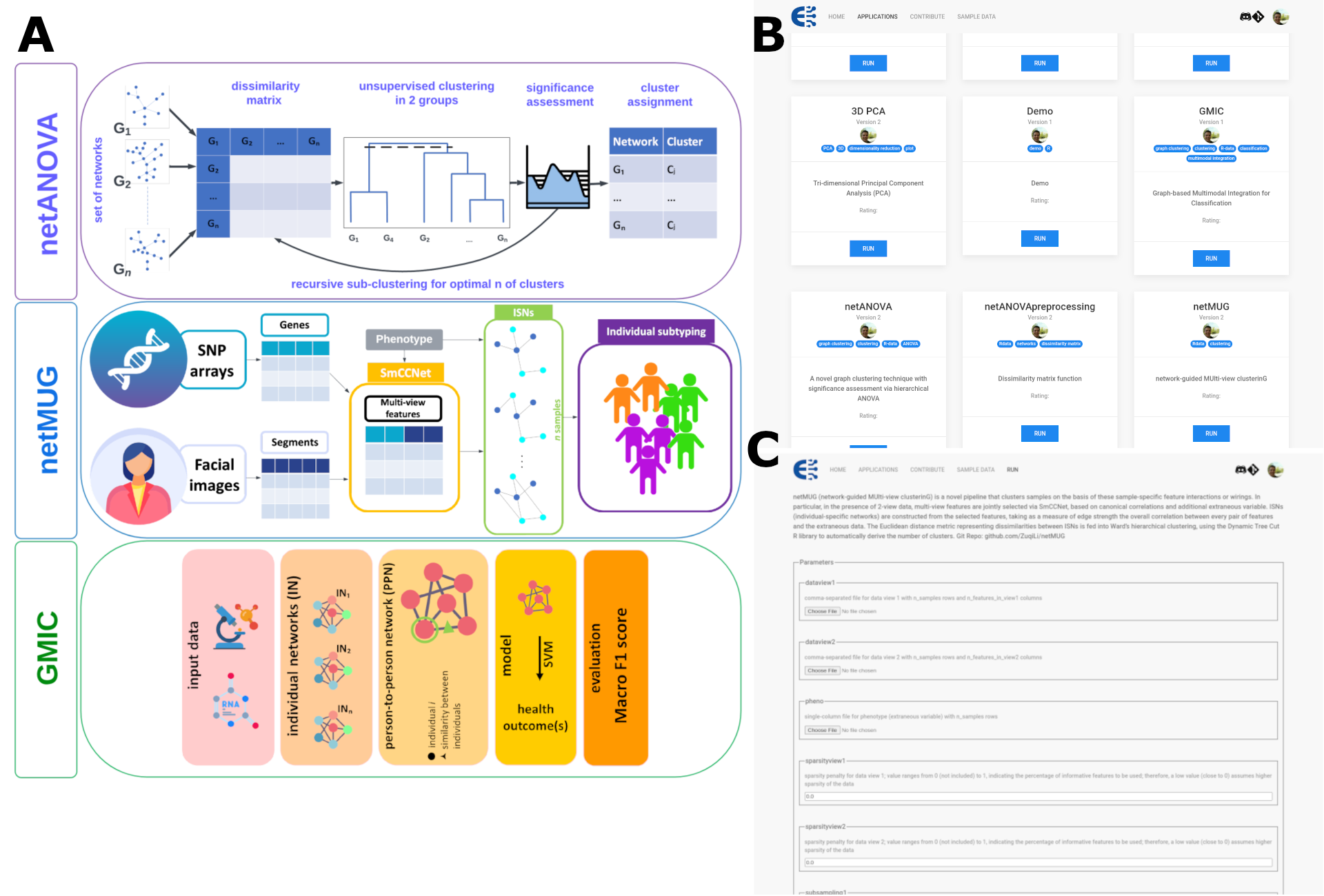}
  \caption{\textbf{A}: A graphical description of the workflow of 3 applications currently available on JINet, adapted from netANOVA~\cite{duroux2023netanova}, netMUG~\cite{li2023netmug}, and GMIC~\cite{duroux2023graph} . \textbf{B} A screenshot from JINet displaying an index of applications.  \textbf{C} Running an application is as simple as graphically selecting the parameters to run with them pressing the ``Run'' button.}
  \label{fig:02}
\end{figure}
\subsection*{User Interface}
JINet is designed with user accessibility and functionality in mind. The interface consists of several key pages that guide the user through the various features and applications available on the platform.

\paragraph{General Information Page:} the initial landing page provides general information about JINet, offering an overview of its capabilities and purpose. This page serves as a starting point for users to familiarize themselves with the platform.

\paragraph{Application Selection Page:} this page lists all available applications that users can choose from. The user may filter applications by tag or by text search. Each application is displayed with a brief description to help users decide which analysis to run. Notably, running an application does not require the user to log in, ensuring ease of access and use.

Upon selecting an application, the user is taken to the \textit{Run} page for that application, where the following is displayed:

\begin{itemize}
    \item \textbf{application name:} and the version that is running,
    \item \textbf{a description:} of the application itself in a longer, more detailed form than in the application list page (this can include a link to the application source code),
    \item \textbf{parameter inputs:} all required parameters for the application, accompanied by concise descriptions;
    \item \textbf{data selection:} users can select local data on which the analysis will be performed.
\end{itemize}
Once all parameters are configured, and data is selected, the user can run the application by clicking on the "Run" button.

\paragraph{Results Retrieval and Sharing:} after the application completes execution, users can view the results on the web page. Where an application makes its output available as a file, a download button will be displayed, allowing the user to save the output to a file outside the browser sandbox. Additionally, JINet provides an option to share results securely. Shared results are encrypted by the web browser before being transmitted to the JINet Server. Once stored on the server the user can share a link with others. The recipient can decrypt the result using a passphrase set by the user who generated the result. JINet requires users to log in in order to share results to minimize the risk of server storage being misused for malicious purposes.

\paragraph{Data Sharing and Contribution:} JINet supports sample data sharing and developer contributions through dedicated forms:

\begin{itemize}
    \item \textbf{sample data upload:} data owners can share non-sensitive versions of their datasets by logging in and requesting permission. Once granted, they can upload sample data, which becomes available for download by other users;
    \item \textbf{application contributions:} developers can contribute their applications to the platform. This process is similar to the data upload steps, requiring login and a permission request.
\end{itemize}


\section*{Limitations}
\label{limitations}
Limitations can be broken down into those that are fundamental to the design of JINet, and those that can be fixed or mitigated with more time and engineering investment.

\subsection*{Fundamental limitations}
JINet applications run within a web browser process. For instance, this makes JINet unsuitable for distributing computations on a compute cluster. The filesystem is represented in memory on non-Chromium based browsers, so large files may lead to the application being stopped prematurely. Whether this is a fundamental limitation depends on the standardisation of native filesystem access within web browsers.

\subsection*{Temporary limitations}
JINet can still improve security, performance, and usability. There is no in-principle reason the application URL white-list should not be empty (not allowed to access any URL; except the JINet application URL) so with future engineering effort, JINet can become even more resistant to leaking user data. Where applications require extra files in order to function properly, these can be provided by JINet.

The relatively poor performance of numerical compute workloads (see Section~\ref{ref:performance}) should disappear with time as libraries like OpenBLAS and PyTorch~\cite{pytorch2} are compiled for the WASM platform. This performance gap will be further reduced as browsers implement the WebGPU \cite{WebGPU} standard and libraries start making use of it.

There are remaining engineering tasks to make the platform more streamlined and easy-to-use. These include friendly error reporting, user interface improvements for users and application developers. Future updates could extend the supported language runtimes to include Julia and other languages, further broadening its applicability and versatility.


\section*{Applications}

Below (Table\ ref{tab1}) is a comprehensive list of applications currently available on JINet, each accompanied by a brief description and their respective runtime. This list is to be considered provisional and continuously evolving, as new applications are anticipated to be added over time. Additionally, several already prepared applications \cite{lalli2024isn, raimondi2021novel, li2024novel} are pending optimisation to address the technical limitations discussed in the main paper, section \textit{Limitations}. This dynamic nature ensures that JINet remains at the forefront of data analysis and sharing capabilities, adapting to the latest advancements and user needs. 

\begin{longtable}{@{\extracolsep{\fill}}lp{0.6\textwidth}p{0.1\textwidth}@{}}
\caption{Details of currently available applications at the time of writing.} \label{tab1} \\
\toprule
\multicolumn{3}{@{}c@{}}{\vspace{0.3em} \Large\centering Application Details \vspace{0.3em}} \\
\addlinespace[0.3em]
\textit{Name} & \textit{Brief Description} & \textit{Runtime} \\
\midrule
\endfirsthead

\multicolumn{3}{c}%
{{\tablename\ \thesupptable{} -- Continued from previous page}} \\
\toprule
\multicolumn{3}{@{}c@{}}{\vspace{0.3em} \Large\centering Application Details \vspace{0.3em}} \\
\addlinespace[0.3em]
\textit{Name} & \textit{Brief Description} & \textit{Runtime} \\
\midrule
\endhead

\midrule
\multicolumn{3}{r}{{Continued on next page}} \\
\bottomrule
\endfoot

\bottomrule
\endlastfoot

netANOVApreprocessing & The function generates a dissimilarity matrix representing the pairwise dissimilarities between networks, saved in "output.txt". The file can be downloaded. More in \cite{duroux2023netanova}. & R \\
& & \\
netANOVA & Requires a dissimilarity matrix derived from pairwise dissimilarities between objects (e.g., networks). Uses hierarchical clustering to identify similar object classes and generate network group memberships. Provides detailed statistics and p-values that can be downloaded. More in \cite{duroux2023netanova} & R \\
& & \\
netMUG & netMUG (network-guided MUlti-view clusterinG) clusters samples based on specific feature interactions. Uses SmCCNet for 2-view data feature selection via canonical correlations and an extraneous variable. Constructs individual-specific networks (ISNs) from selected features, measuring edge strength by overall correlation with extraneous data. Applies Ward’s hierarchical clustering with Dynamic Tree Cut R library to determine cluster number. Provides downloadable clusters. More in \cite{li2023netmug}. & R \\
& & \\
GMIC & GMIC (Graph-based Multimodal Integration for Classification) aims to perform supervised learning (classification) by integrating multiple data modalities. It takes raw data with samples in rows and features in columns. Data modalities are integrated through individual networks where nodes and edges are specific to each individual. The methodology follows: (1) creation of individual networks using the node product approach; (2) computation of similarity between individuals based on these graphs at both the node and edge levels. Node-level similarity is computed using Spearman correlation, while edge-level similarity is calculated using the edge difference distance; (3) integration of the similarity matrices by averaging them; (4) training of a Support Vector Machine (SVM) model on the integrated similarity matrices; (5) utilization of the trained model to predict group labels for new data. More in \cite{duroux2023graph}. & R \\
& & \\
2D, 3D PCA & Dimensionality ML (Machine Learning) method for reducing the number of variables in a data set while preserving as much information as possible. Provides an HTML string displayed directly in the user interface that can be downloaded. & Python \\
& &  \\
PCA loadings & PCA Loadings is a ML method designed to reduce the dimensionality of a dataset while retaining maximal information content. It transforms the original variables into a new set of orthogonal components called loadings. These loadings are computed as the eigenvectors of the covariance matrix, scaled by the square root of their corresponding eigenvalues. Provides an HTML string displayed directly in the user interface that can be downloaded. & Python \\
& & \\
2D tSNE & T-distributed stochastic neighbour embedding (t-SNE) is a technique that helps users visualize high-dimensional data sets. It converts similarities between data points to joint probabilities and tries to minimise the Kullback-Leibler divergence between the joint probabilities of the low-dimensional embedding and the high-dimensional data.  Provides an HTML string displayed directly in the user interface that can be downloaded. & Python \\
& & \\
2D UMAP & Uniform Manifold Approximation and Projection (UMAP) is a dimension reduction technique that can be used for visualisation similarly to t-SNE, but also for general non-linear dimension reduction. Provides an HTML string displayed directly in the user interface that can be downloaded. & Python \\
& & \\
Scatter Matrix & A scatter plot matrix is a grid (or matrix) of scatter plots used to visualise bivariate relationships between combinations of variables. Each scatter plot in the matrix visualises the relationship between a pair of variables, allowing many relationships to be explored in one chart. Provides an HTML string displayed directly in the user interface that can be downloaded. & Python\\
& & \\
Scatter Marginals & Scatter Plot with Marginal Histograms is a joint distribution plot with the marginal distributions of two selected variables. Provides an HTML string displayed directly in the user interface that can be downloaded. & Python\\
& & \\
ROC binary, multiclass & The Receiver Operator Characteristic (ROC) curve is an evaluation metric for binary and multiclass classification problems. It is a probability curve that plots the TPR against FPR at various threshold values and essentially separates the \textit{signal} from the \textit{noise}. Provides an HTML string displayed directly in the user interface that can be downloaded. & Python \\
& & \\
PR binary, multiclass & Precision-Recall is a useful measure of success of prediction when the classes are very imbalanced. In information retrieval, precision is a measure of result relevancy, while recall is a measure of how many truly relevant results are returned. The precision-recall curve shows the tradeoff between precision and recall for different thresholds. Provides an HTML string displayed directly in the user interface that can be downloaded. & Python\\
& & \\
LR preliminary plots & Linear regression analysis is used to predict the value of a variable based on the value of another variable. The variable you want to predict is called the dependent variable. The variable you are using to predict the other variable's value is called the independent variable. Provides an HTML string displayed directly in the user interface that can be downloaded. & Python\\
& & \\
Demo & Application demonstrating arbitrary HTML output. Provides an HTML string displayed directly in the user interface that can be downloaded. & R \\
\end{longtable}

\section*{Performance}
\label{ref:performance}
Performance was evaluated on optimized numerical workloads and pure language evaluation in R and in Python. Evaluation of R and Python code is similar between native and WASM interpreters. Neither WASM runtime (R or Python) can offload numeric workloads to a high-performance library like OpenBLAS yet. This is where the major performance differences lie.

The machine used to run these benchmarks is an Intel Core i5-8265U laptop with 16GiB of RAM. The WASM benchmarks were run in JINet running on Chromium version $125.0.6422.60$.

\begin{figure}[htbp]
  \centering
  \includegraphics[width=\textwidth]{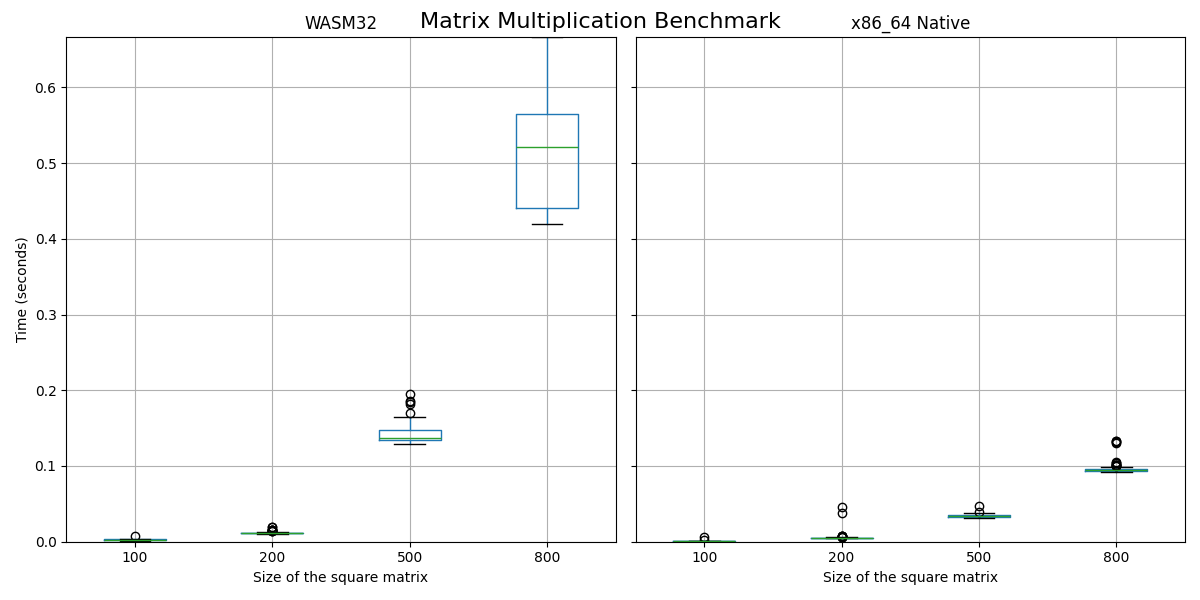}
  \vspace*{-25pt}
  \caption{Matrix multiplication in the WASM R interpreter (left) vs. native matrix multiplication with OpenBLAS (right). Each iteration of the benchmark generates 2 random square matrices which are then multiplied with the \texttt{\%*\%} operator. This benchmark is run 100 times for each matrix size.}
  \label{fig:R-matmul}
  \vspace*{-5pt}
\end{figure}

The native R interpreter uses the OpenBLAS numerical library to
accelerate matrix multiplication. This is not yet available on the WASM interpreter so this benchmark illustrates a significant reduction in performance running numerical workloads on WASM compared to native (see Figure \ref{fig:R-matmul}). The native benchmark was run with \texttt{OPENBLAS\_NUM\_THREADS=1} to ensure a fair comparison. We expect that this performance gap will dramatically reduce when the WASM R interpreter gains access to OpenBLAS.

\begin{figure}[htbp]
  \centering
  \includegraphics[width=\textwidth]{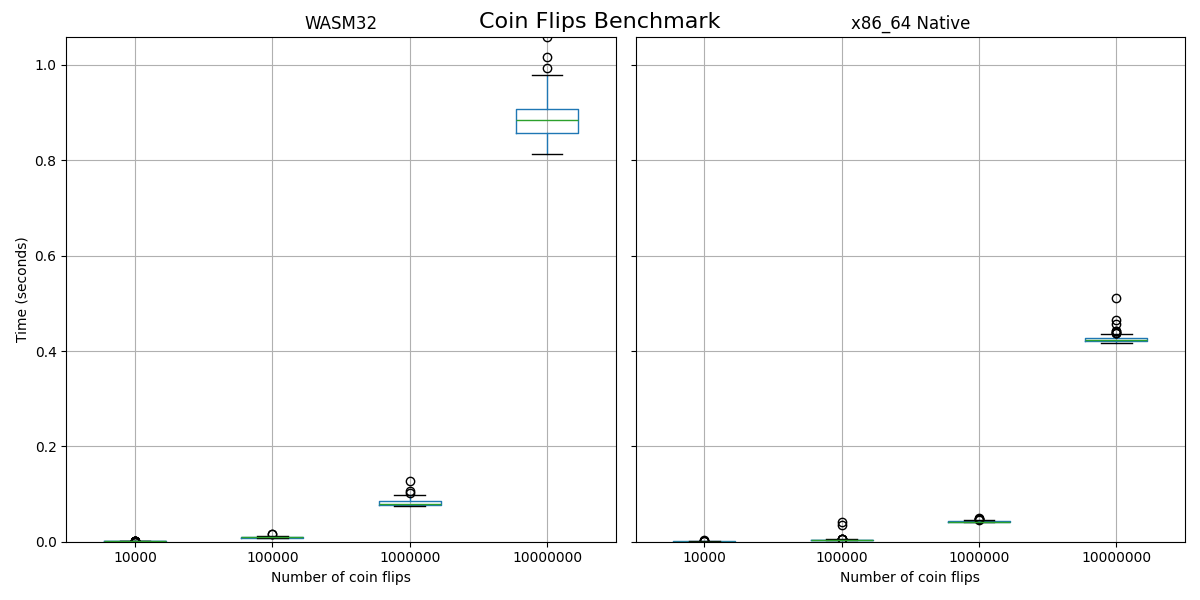}
  \vspace*{-25pt}
  \caption{Simulated coin flips in the WASM R interpreter (left) vs. native R interpreter (right). Each iteration of the benchmark generates a large vector containing the string ``\texttt{H}'' or ``\texttt{T}'' selected randomly. The number of ``\texttt{H}'' strings are then counted. This benchmark is run 100 times for each vector size.}
  \label{fig:R-coin-flips}
  \vspace*{-5pt}
\end{figure}

Native R code evaluation performance was evaluated by sampling a large number of coin flips and counting how many are heads (see Figure \ref{fig:R-coin-flips}). This workload is not offloaded to OpenBLAS so is largely reflective of the performance difference between native R and WASM R.

\begin{figure}[htbp]
  \centering
  \includegraphics[width=\textwidth]{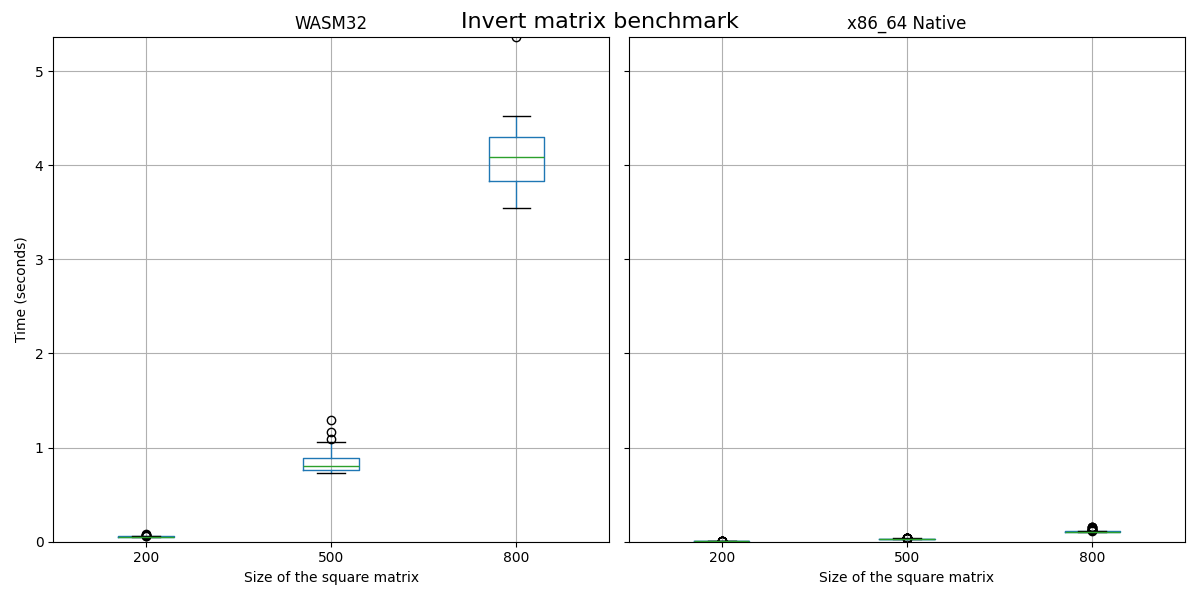}
  \vspace*{-25pt}
  \caption{Computation of an inverse matrix from a randomly generated square matrix in the WASM Python interpreter (left) vs. native Python interpreter with OpenBLAS (right). Each iteration of the benchmark generates a random square matrix which is then inverted. This benchmark is run 100 times for each matrix size.}
  \label{fig:Python-matrix-inverse}
  \vspace*{-5pt}
\end{figure}

Similar performance comparisons were performed for the Python interpreter. The Python NumPy~\cite{numpy} library uses OpenBLAS to compute matrix inverses. As with R, at the time of writing, Pyodide does not build Numpy with OpenBLAS so performance is much worse than native Python with OpenBLAS. The native benchmark was run with \texttt{OPENBLAS\_NUM\_THREADS=1} to ensure a fair comparison. The Numpy library was used to invert a randomly generated square matrix (see Figure \ref{fig:Python-matrix-inverse}).

\begin{figure}[htbp]
  \centering
  \includegraphics[width=\textwidth]{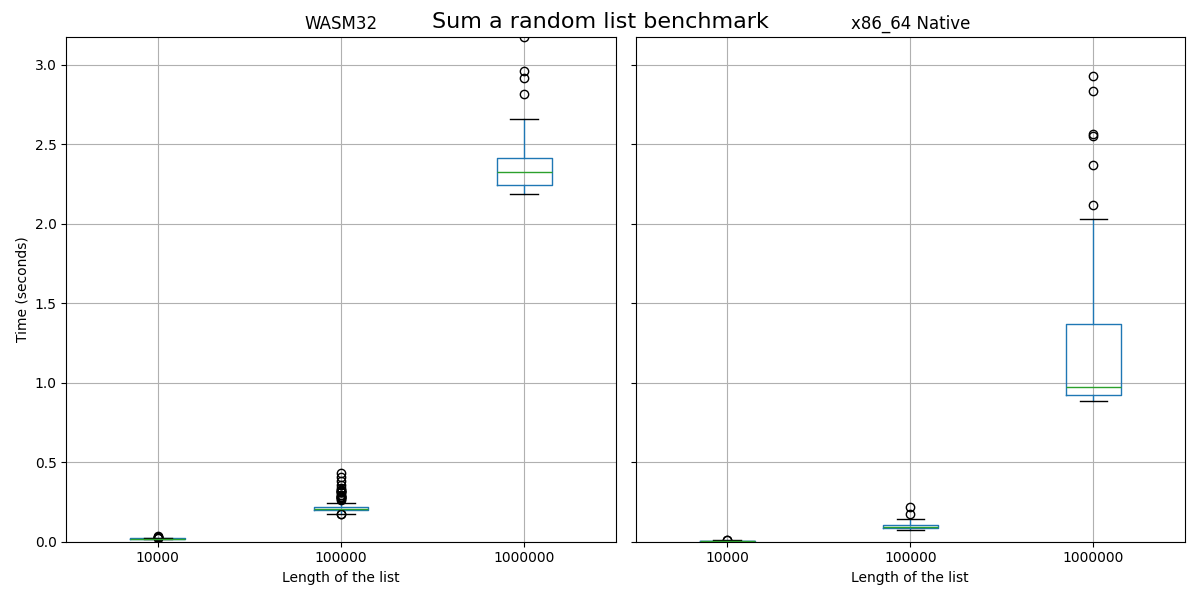}
  \vspace*{-25pt}
  \caption{Summing a list of numbers in the WASM Python interpreter (left) vs. native Python interpreter (right). Each iteration of the benchmark generates a large list containing random numbers which is then passed to the built-in \texttt{sum} function. This benchmark is run 100 times for each list size.}
  \label{fig:Python-list-sum}
  \vspace*{-5pt}
\end{figure}

Finally, Python evaluation performance was measured by summing the contents of a random numerical list of varying size (see Figure \ref{fig:Python-list-sum}).

\pagebreak

\bibliography{document}

\end{document}